\RequirePackage{fix-cm}
\documentclass[twocolumn]{svjour3}          
\smartqed  
\usepackage[usenames,dvipsnames]{xcolor}
\usepackage{natbib}
\usepackage{graphicx}
\usepackage{amsmath}
\usepackage[normalem]{ulem} 
\usepackage{lineno}

\graphicspath{{./Figs/}}

\newcommand{\mg}{\color{black}}

\newcommand{\x}{\mathbf{x}}
\newcommand{\X}{\mathbf{X}}
\newcommand{\y}{\mathbf{y}}

\newcommand{\Y}{\mathbf{Y}}
\newcommand{\vv}{\mathbf{v}}
\newcommand{\w}{\mathbf{w}}

\newcommand{\vF}{\mathbf{F}}

\newcommand{\Hs}{\textrm{H}}
\newcommand{\M}{\textrm{M}}
\newcommand{\MM}{\mathcal{M}}

\def\y{{\mathbf y}}
\def\bP{{\mathbf P}}
\def\bR{{\mathbf R}}
\def\bH{{\mathbf H}}
\def\b0{{\mathbf 0}}
\def\bSigma{{\mathbf \Sigma}}
\def\({\left(}
\def\){\right)}

\journalname{{\mg Climatic Change}}
\begin{document}

\title{DADA: Data Assimilation for the Detection and\\ Attribution of Weather- and Climate-related Events}



\titlerunning{{\mg DADA: Data Assimilation for Detection \& Attribution}} 

\author{A. Hannart  \and A. Carrassi \and M. Bocquet \and M. Ghil \and P. Naveau \and M. Pulido \and J. Ruiz \and P. Tandeo}


\institute{A. Hannart \at
              IFAECI, CNRS-CONICET-UBA\\
              Pab. II, piso 2, Ciudad Universitaria\\
              1428 Buenos Aires, Argentina\\
              Tel.: +5411-4787-2693\\
              Fax: +5411-4788-3572\\
              \email{alexis.hannart@cima.fcen.uba.ar}           
 \and
A. Carrassi \at Mohn-Sverdrup Center, Nansen Environmental and Remote Sensing Center, Bergen, Norway\\
           \and
M. Bocquet \at CEREA, \'Ecole des Ponts, Marne-la-Vall\'ee, France\\
 \and
M. Ghil \at Ecole Normale Sup\'erieure, Paris, France\\
University of California, Los Angeles, USA\\
 \and
 P. Naveau \at LSCE, CNRS, Gif-sur-Yvette, France\\
 \and
 M. Pulido \at Dept. of Physics, Universidad Nacional del Nordeste, Corrientes, Argentina\\
 \and
 J. Ruiz \at IFAECI, CNRS/CONICET/UBA, Buenos Aires, Argentina\\
 \and
 P. Tandeo \at T\'el\'ecom Bretagne, Brest, France
}

\date{Received: date / Accepted: date}

\maketitle

\begin{abstract}
We describe a new approach allowing for 
systematic causal attribution of weather and climate-related events, in near-real time. The method is purposely designed to 
facilitate its implementation 
at meteorological centers by 
relying on data treatments that are routinely performed when numerically forecasting the weather. 
Namely, we show that causal attribution can be obtained as a by-product of so-called \textit{data assimilation} procedures that are run on a daily basis to update the meteorological model with new atmospheric observations; hence, the proposed methodology can take advantage of the powerful computational and observational capacity of weather forecasting centers. We explain the theoretical rationale of this approach and sketch the most prominent features of a ``data assimilation based detection and attribution'' (DADA) procedure. The proposal is illustrated in the context of the classical three-variable Lorenz model with additional forcing. Several theoretical and practical 
research questions that need to be addressed to make the proposal readily operational within weather forecasting centers are finally laid out.
\keywords{{\mg Event attribution \and Data assimilation \and Causality theory \and Modified Lorenz model}}
\end{abstract}

\section{Background and motivation}
\label{sec:intro}

A significant and growing part of climate research studies the causal links between climate forcings and observed responses. This part has been consolidated into a research topic known as detection and attribution (D\&A). The D\&A community has increasingly been faced with the challenge of generating causal information about episodes of extreme weather or unusual climate conditions. This challenge arises from the needs for public dissemination, litigation in a legal context, adaptation to climate change or simply improvement of the science associated with these events \citep{Stott13}.

The 
approach 
widely used so far to in D\&A was introduced one decade ago by M.R. Allen and colleagues \citep{All03,Stone05} and it originates from best practices in epidemiology \citep{GreenRoth98}. In this approach, one evaluates the extent to which a given external climate forcing --- such as solar irradiation, greenhouse gas (GHG) emissions, ozone or aerosol concentrations --- has changed the probability of occurrence of an event of interest. 

For this purpose, one compares the probability of occurrence of said event in an ensemble of model simulations representing the observed climatic conditions, which simulates the actual occurrence probability in the real world, 
with the occurrence probability of the same event in a parallel ensemble of model simulations, which represent an alternative world. The 
former world is referred to as {\em factual,} the latter as {\em counterfactual}: 
it is the latter that might have occurred had the forcing of interest been absent.

Denoting by $p_1$ and $p_0$ the probabilities of the event occurring in the factual world and in the counterfactual world respectively, the so-called fraction of attributable risk (FAR) is then defined as FAR$=1-p_0/p_1$.
The FAR has long been interpreted as the fraction of the likelihood of an event which is attributable to the external forcing. Over the past decade, most causal claims have been following from the FAR and its uncertainty, resulting in statements such as ``\textit{It is very likely that over half the risk of European summer temperature anomalies exceeding a threshold of 1.6$^\circ$C is attributable to human influence.}'' \citep{Stott04}. 

\cite{HPONG14} have recently shown that, under realistic assumptions, the FAR may also be interpreted as the so-called \textit{probability of necessary causation} (PN) associated --- in a complete and self-consistent theory of causality \citep{Pearl00} --- with the causal link between the forcing and the event. The FAR thus corresponds to only one of the two facets of causality in such a theory, while the \textit{probability of sufficient causation} (PS) is its second facet.\\ 
In this setting,
\begin{linenomath*}\begin{subequations}\label{proba}
\begin{align}
& \textrm{PN} =1 - \frac{p_0}{p_1}, \label{proba1} \\
& {\textrm{PS}} =1- \frac{1-p_1}{1-p_0}\,, \label{proba2} \\
& \textrm{PNS} = p_1-p_0\,, \label{proba3} 
\end{align}
\end{subequations}\end{linenomath*}
where PNS is the \textit{probability of necessary and sufficient causation.}

\cite{Pearl00} provides rigorous definitions of these three concepts, as well as a detailed discussion of their 
meanings and implications. 
It can be seen from Eqs. (\ref{proba}) that causal attribution requires to evaluate the two probabilities, $p_0$ and $p_1$, 
and not just one of them. Doing so 
is, therefore, the central methodological question of D\&A for weather and climate-related events. 

So far, most case studies have 
used large ensembles of climate model simulations in order to estimate $p_1$ and $p_0$ based on a variety of methods, in particular based on statistical extreme value theory (EVT). 
However, this general approach has a very high computational cost and is difficult to implement in a timely and systematic way. As recognized by \cite{Stott13}, this remains an open 
problem: 
``the overarching challenge for the community is to move beyond research-mode case studies and to develop systems that can deliver regular, reliable and timely assessments in the aftermath of notable weather and climate-related events, typically in the weeks or months following (and not many years later as is the case with some research-mode studies)''. For instance, the \texttt{weather@home} system \citep{Mass14}, or the system proposed by \cite{Christ13}, aim at meeting those requirements within the conventional ensemble-based approach. Ongoing research aiming towards the development of such a system also include the CASCADE project (Calibrated and Systematic Characterization, Attribution and Detection of Extremes, U.S. Department of Energy, Regional and Global Climate Modeling program).

The purpose of this article is to introduce a new methodological 
approach that addresses the latter overarching operational challenge. Our proposal relies on a class of powerful statistical 
methods for interfacing high-dimensional models with large observational datasets. This class of methods originates from the field of weather forecasting and is referred to as \textit{data assimilation} (DA) \citep{BGK81,GMR91,OT97}. 

{\mg Section~\ref{sec:method}}  explains the rationale of the approach proposed herein, presents a brief overview of DA, and 
outlines the most prominent technical features of a ``data assimilation--based detection and attribution'' (DADA) approach. {\mg Section~\ref{sec:L63}} illustrates the proposal by implementing it on 
a version of the classical Lorenz convection model \cite[L63 hereafter]{Lor63} subject to an additional constant force. {\mg Finally, in Section~\ref{sec:concl}, we discuss} the main strengths and limitations of the DADA approach, and highlight several theoretical  and practical 
research questions that need to be addressed to make it potentially operational within weather forecasting centers in a near future.


\section{Method description}
\label{sec:method}

\subsection{General rationale}
\label{ssec:rationale}
The rationale 
{for addressing} causal attribution of climate-related events based on DA 
{concepts and methods can 
be outlined} in three steps. To {do so 
briefly and clearly,} we need to introduce 
{some notation.}

Let $\y_t$ denote the $d$-dimensional vector of observations at discrete times 
{$\{t = 0,1,\ldots,T\}$.} Here, {$\y = \{\y_t: 0 \leq t \leq T\}$} corresponds, for instance, to the full set of all available meteorological observations over a time 
{interval} covering the event of interest, no matter the diversity and source of the data; {typically, the latter include} ground station networks, satellite measurements, {ship data, and so on, cf.~\cite[Preface, Fig.~1]{BGK81} or \cite[Fig.~1]{GMR91}.}  In the present probabilistic D\&A context, the observed trajectory $\y$ is viewed as a realization of a random variable denoted 
$\Y = \{\Y_t: 0 \leq t \leq T\}$, i.e. there exists an $\omega\in\Omega$ such that $\Y(\omega) = \y$ --- where $\Omega$ denotes the sample space of all possible outcomes and encompasses observational error, as well as internal variability. 

{In event attribution studies, it is recognized that defining the \textit{occurrence} of an event, i.e. selecting a subset $\mathcal{F}\subset\Omega$, depends on a rather arbitrary choice.}
Yet 
{this choice} has been shown to greatly affect causal conclusions \citep{HPONG14}. For instance, a 
generic {and fairly loose} event definition is arguably prone to yield a 
{low threshold} of evidence 
{with respect to} both necessary and sufficient causality while, on the other hand, a {tighter and more specific} event definition is prone to yield a stringent 
{threshold for} necessary causality but a reduced one {for} sufficient causality.

Indeed, it is quite intuitive that many different factors should usually be \textit{necessary} to trigger the occurrence of a highly specific event and conversely, that no single factor will ever hold as a \textit{sufficient} explanation thereof. For the class of \textit{unusual} events at stake in D\&A, where {both $p_0$ and $p_1$ 
are very small,} we arguably lean towards specific definitions that inherently result 
{in few sufficient causal factors or none. 
This conclusion immediately follows from Eq.~(1b),}
which yields $\textrm{PS}\simeq 0$ when both $p_0$ and $p_1$ are very small. 

Usually, an event occurrence is defined in D\&A based on an \textit{ad hoc} scalar index $\phi(\Y)$ exceeding a threshold $u$, i.e. $p_i=P(\phi(\Y)\geq u)$; 
{from now on, we associate $i = 0$ with the counterfactual and $i = 1$ with the factual world.} 
While this definition may be {already} quite restrictive for $u$ large, 
it is 
a {defensible} strategy to restrict the event definition even further: 
this may slightly reduce an already negligible $\textrm{PS}$ but in return may potentially increase $\textrm{PN}$ by a greater amount; 
{one thus expects 
to gain more than 
one loses in this trade-off.} In particular, this will be the case if additional features, not accounted for in $\phi(\Y)$, can be identified that {will allow 
one to further discriminate between the two} worlds. 

In any case, a central element of our proposal 
is to follow this strategy in its simplest {possible} form, by using the tightest 
occurrence definition i.e. the singleton $\{ \omega\in\Omega\mid\Y(\omega)=\y\}$. Note that the latter singleton has probability zero in both worlds because the {probability density function (PDF) $f(\Y(\omega))$} of $\Y$ can be assumed, {in general, to be} continuous, i.e. 
to contain no singular $\delta$-functions. 

Consider, however, the paradox that arises from taking the limit $h \to 0$ for the set $\{ \omega\in\Omega\mid \Vert\Y(\omega)-\y\Vert\leq h\}$. {This set} has non-zero probability for $h$ arbitrarily small but positive 
while, in the limit, 
\begin{linenomath*}\begin{equation}
\label{proba_sing}
\textrm{PN} =1 - \frac{f_0(\y)}{f_1(\y)},\quad{\textrm{PS}} =0, 
\end{equation}\end{linenomath*}
where $f_i$ denotes the PDF of $\Y$ in world $i$. 
Equation (\ref{proba_sing}) thus shows that, while the probabilities of occurrence of our singleton event in both worlds are null, its associated probability of necessary causation is 
{\mg still} positive --- but its probability of sufficient causation is always zero. Our proposal thus intentionally sacrifices 
evidence {of sufficiency,} in the hope of maximizing the evidence of necessity.

{Our betting on the singleton set is thus} justifiable {already} based 
on the above theoretical considerations. {This choice, 
moreover, is motivated by having} a highly simplifying implication from a practical standpoint. 
Evaluating the PDF of $\Y$ at a single point $\Y=\y$ is indeed, under many circumstances, considerably 
{easier} than evaluating the probability $P(\phi(\Y)\geq u)$ required in the conventional approach. 

To illustrate this point, let $\Y$ be for instance a {$d$-}variate autoregressive process defined by $\Y_{t+1}=\mathbf{A}\Y_t +\w_t$, where $\w_t$ is an i.i.d. noise having known PDF {$g(\cdot)$} and where $\mathbf{A}$ has the 
{usual properties that insure stationarity 
\citep{Gard04}.} We {then} have:
\begin{linenomath*}\begin{subequations}
\label{AR1}
\begin{align}
&f(\y) = \prod_{t=1}^{T} g(\y_{t}-\mathbf{A}\y_{t-1})\pi(\y_0)\,, \label{AR1a} \\
\begin{split} P(\phi(\Y)\geq u) &= \int_{\phi(\y)\geq u} \prod_{t=1}^{T} g(\y_{t}-\mathbf{A}\y_{t-1}) \\
&\times\pi(\y_0) \textrm{d}y_{1,0} \ldots \textrm{d}y_{d,0} \ldots \textrm{d}y_{d,T}\,, \end{split} \label{AR1b} 
\end{align}
\end{subequations}\end{linenomath*}
with {$\pi(\cdot)$} the prior PDF on the initial 
{state $\Y_0$. Equation~\eqref{AR1a} shows} that $f(\y)$ can be easily computed using a {closed-form} expression, while 
$P(\phi(\Y)\geq u)$ {in Eq.~\eqref{AR1b} 
is an integral on $d\times T+1$ dimensions which 
must instead be evaluated 
by using, for instance, a {computationally quite costly} Monte-Carlo (MC) simulation. 

{Figure~\ref{pdfex}}
illustrates this situation by showing the details of the latter MC evaluation {for a scalar AR(1) process (panel $a$,} when based on a standard EVT 
{application,} as well as its associated 
{accuracy (panels $b$ and $c$), and the computational} cost {as} the MC sample size {$n$ varies (panel $d$); the latter cost is much larger than the one of applying the DADA approach.}  
This simple example 
{confirms the large} computational discrepancy {between the two approaches.} 

{The reason for the discrepancy} is quite simple: evaluating the conventional probability requires integrating a PDF over a predefined domain, instead of a one-off evaluation at a single point. Because both the domain of integration and the PDF may have potentially complex shapes, 
{one cannot expect, in general, that the requisite integral be 
amenable to analytical treatment.} 
Hence numerical integration is the default option: no matter 
{how efficient an integration scheme one applies, it will require} evaluating the PDF at many points and is thus as many times more costly computationally than {just} evaluating $f(\y)$ {at a single point.}

This being said, 
{it is not} always straightforward to obtain {the PDF of $\Y$.} 
This is the case, for instance, for the wide class of statistical models referred to as \textit{Hidden Markov Models} {(HMMs); 
in fact, HMMs [e.g., \cite[and references therein]{Ihler07}] are}
often relevant in the present context to describe $\Y$. 


More precisely, assume that the event of interest can be represented by a large numerical model which $N$-dimensional state vector at time $t$ is denoted $\X_t$. The dynamics of the state vector is given by:
\begin{linenomath*}\begin{equation}
\label{dyn}
\X_{t+1} = {\M(\X_{t},\vF_{t})} + \vv_{t}\,,
\end{equation}\end{linenomath*}
where $\M$ is the model operator, $\vv_t$ is a stochastic term representing modeling error, and {$\vF_t$} is a known, prescribed forcing that is external to the model. In the present context, {it is precisely} the forcing term {$\vF = (\vF_t)_{t=0}^T$ that} is under causal scrutiny. Further, assume that our observations $\Y_t$ can be mapped to the state vector $\X_t$ at any time $t$, i.e.
\begin{linenomath*}\begin{equation}
\label{obs}
\Y_t=\Hs(\X_t)+\w_t
\end{equation}\end{linenomath*}
where $\Hs$ is the so-called {observation or} forward operator and $\w_t$ is a stochastic term representing observational error. 

Denoting {by $\vF^{(i)}$} the value of the forcing in the world $i$, using the shorthand $\M_i(x_{t}) 
= \M(\x_{t},\vF_{t}^{(i)})$ and denoting by $\MM_i$ the HMM associated {with} $\Hs$ and $\M_i$, the problem of interest here is thus to derive:
\begin{linenomath*}\begin{equation}
\label{p1p0}
f_0(\y)=f(\y\mid\MM_0) {\quad \textrm{and} \quad f_1(\y)=f(\y\mid\MM_1)\,,}
\end{equation}\end{linenomath*}
where $f_0(\y)$ and $f_1(\y)$ 
{should be interpreted} as the  likelihoods of the observation $\y$ 
{in} the counterfactual and factual models, {respectively.} 

Finally getting to our point, 
{one can view} DA methods 
as a class of inference methods designed for the above HMM setting. {Actually, \cite{Ihler07} already formulated both DA and HMMs within the broader class of graphical models for statistical inference.}

While inferring the unknown state vector trajectory $\X$, 
{given} the observed trajectory $\y$, is clearly the main focus of DA, the likelihood $f(\y)$ can also be obtained as a side product thereof, as {we will 
immediately clarify below.} 
Therefore, 
with DA 
{able to} derive the two likelihoods $f_0(\y)$ and $f_1(\y)$, and the latter two being the keys to causal attribution in our approach,
one should be capable of moving towards near-real-time, systematic causal attribution of {weather-} and climate-related events. 

\subsection{Brief overview of {data assimilation}} 
\label{ssec:DA}

DA 
{was initially developed} in the context of numerical weather forecasting, in order to initialize the model's state variables $\X$ based on observations $\y$ that are {incomplete, diverse in nature, unevenly distributed in space and time,} do not necessarily match the model's state variables, and are contaminated by measurement error {\citep{BGK81,OT97}.} Over the past decades, those methods have grown out of their original application field to reach a wide variety of topics in geophysics such as oceanography {\citep{GMR91}}, atmospheric chemistry, {geomagnetism, hydrology, and space physics, among many other areas \citep{Rob06,Cos10,Kond11,Boc12,Mar14}.}

{DA is already playing an increasing role in the climate sciences, 
having being applied,} for instance, 
to initialize a climate model for seasonal or decadal prediction {\citep{Bal09},} to constrain a climate model's parameters {\citep{Kond08,Rui13},} to infer carbon cycle fluxes from atmospheric concentrations \citep{Che13}, or to reconstruct paleoclimatic fields out of sparse and indirect observations {\citep{Bhe12,Roque14}.} In the context of D\&A, 
\citet{Lee08} {actually tested a DA-like approach to 
include 
the effects of the various forcings over the last millennium, in addition to other paleoclimate proxy data, in combined climate reconstruction and detection analysis.} The present work thus 
{follows and further strengthens a} general trend {in climate studies.}

Methodologically speaking, DA 
{methods} are traditionally grouped into two categories: sequential and variational {\citep[and references therein]{Ide97}.} In the sequential 
approach 
{\citep{Ghil81},} the state estimate and 
a suitable estimate of the associated error {covariance matrix are} propagated in time until new observations become available and are used {to update} the state estimate. 
In practice, the evolution of the system of interest is retrieved {--- like in earlier, typically much smaller-dimensional applications \citep{Kal60,Jaz70,Gelb74} ---} through a sequence of prediction and analysis steps. 
In the variational approach, on the other hand, {one seeks} 
the system trajectory 
{that} best fits all the observations distributed within a given {time} interval {\citep{LDT86,Ide97,Boc12}.} 
Here, we 
concentrate on the sequential approach, 
{but the two approaches are complementary and the choice of method depends on the specifics of the problem at hand \citep{GMR91,Ide97,OT97}.}

{Abundant} literature is available 
on DA and 
{on Kalman-type filters. 
\cite{Kal60} first presented the solution in discrete time} 
for the case 
{in which both the dynamic evolution operator $\M$ in Eq. \ref{dyn} and the observation operator $\Hs$ in Eq. \ref{obs} are linear, and the errors are Gaussian.}
Under these assumptions, the state-estimation problem for the system given 
{by Eqs. (\ref{obs}, 
\ref{dyn})} has an exact solution given by the following 
{sequential Kalman filter (KF)} equations:
\begin{linenomath*}\begin{subequations}\label{KF}
\begin{align}
& \x^a_t = \x^f_t +\mathbf K(\y_t-\mathbf H \x^f_t) \,, \label{KF1} \\
& \mathbf P^a_{t} = (\mathbf I -\mathbf K\mathbf H) \mathbf P^f_{t} \,, \label{KF2} \\
& \x^f_{t+1} = \mathbf M \x^a_t \,, \label{KF3} \\
&  \mathbf P^f_{t+1} = \mathbf M \mathbf P^a_{t}\mathbf M' + \mathbf Q  \,. \label{KF4} 
\end{align}
\end{subequations}\end{linenomath*}
where $'$ denotes the transpose operation. Here 
Eqs.~\eqref{KF1} and \eqref{KF2} are referred to as the analysis step 
and denoted by a superscript $a$, while the 
forecast step 
is given by Eqs.~\eqref{KF3} and \eqref{KF4}, and is denoted by a superscript $f$ \citep{Ide97}.
{The vector $\x^a_t$ and the matrix} $\mathbf P^a_t$ are the mean and covariance of $\X_t$ conditional on $(\Y_1,...,\Y_t)=(\y_1,...,\y_t)$; $\mathbf K=\mathbf P^f_{t}\mathbf H'(\mathbf H\mathbf P^f_{t}\mathbf H'+\mathbf R)^{-1}$ is the so-called Kalman gain matrix; {while} $\mathbf Q$ and $\mathbf R$ are the covariances associated 
{with} $\vv_t$ and $\w_t$, respectively. 
{Following \cite{Wiener}, one distinguishes between {\em filtering}, in which}
$\x^a_t$ and $\mathbf P^a_t$ are conditioned {only on} the previous and current observations $({\y_0,}...,\y_t)$, {and {\em smoothing}, 
in which they are conditioned} on the entire sequence, {$0 \leq t \leq T$. Furthermore, the 
sequential algorithm} needs to be initialized at time $t=0$ with $\x_0^{f}$ and $\mathbf{P}_0^{f}$, which thus represent the a priori mean and covariance of $\X_0$, respectively, and 
{have to be} prescribed by the user. 

{The} likelihood function $f(\y)$, which is of primary importance 
{for DADA,} also has an exact expression under the 
{above linearity and Gaussianity} assumptions {\citep{Tand14},}  
given by:
\begin{linenomath*}\begin{equation}
\label{llkKF}
\begin{split}
f(\y) = &\prod_{t=0}^{T} \,\,(2\pi)^{-\frac{d}{2}}\vert \mathbf{\Sigma}_t\vert^{-\frac{1}{2}}\\
&\times\exp\left\{-\frac{1}{2}(\y_{t}-\mathbf H\x_t^{f})'\mathbf{\Sigma}_t^{-1}(\y_{t}-\mathbf H\x_t^{f})\right\} \,,\end{split}
\end{equation}\end{linenomath*}
with $\mathbf{\Sigma}_t = \mathbf{H}\mathbf{P}_t^{f}\mathbf{H}'+\mathbf{R}$. 
The proof of {\mg Eq.~\eqref{llkKF}} is 
{provided in the} Appendix, 
{\mg and $f(\y)$ is typically computed by taking the logarithm of this equation to turn the product on the right-hand side into a sum.}

It follows from the above that, once the observations $\y_t$ have been assimilated on the interval {$0 \leq t \leq T$, the 
necessary ingredients  $\x^f_t$ and $\mathbf P^f_t$ in Eq. \ref{llkKF}}
are available and 
{thus calculating $f(\y)$ 
is both straightforward and computationally inexpensive. 
The fundamental connections between this calculation, the HMM context, and} Bayes theorem {are further clarified in the} Appendix. 


{Many difficulties arise in applying the simple ideas outlined here to geophysical models, 
which are typically nonlinear, have non-Gaussian errors and are huge in size \citep{GMR91}. Most of these difficulties have been addressed by improving both sequential and variational methods in several ingenious ways \citep{BPW10,Kond11}.}

In particular, 
the Ensemble Kalman Filter 
{(EnKF; Evensen, 2003)
--- in which the uncertainty propagation is 
evaluated by using a finite-size ensemble of trajectories ---
is} now operational in numerical weather and oceanic prediction centers worldwide; see e.g. \cite{Sak12,Hou14}. 
The EnKF is a convenient {approximate} solution to the filtering problem in a nonlinear, {large-dimensional} context. 
{We simply note here that} it can also be applied to obtain an approximation of the likelihood $f(\y)$ by {substituting the approximate sequence $\{ (\hat{\x}^f_t, \hat{\mathbf P}^f_t): t = 0, \ldots, T \}$ that the EnKF produces into Eq. \ref{llkKF}.}
This strategy is illustrated immediately below in the context of the L63 
{convection model subject to an additional constant force.} \\

\section{Implementation within the {modified} L63 model}
\label{sec:L63}
\subsection{{The modified model and its two worlds}}
\label{ssec:modified}

{A simple modification \citep{Pal99} of the L63 system \citep{Lor63}} has been extensively used for the purpose of illustrating methodological developments in both DA and D\&A {[e.g. \citep{CV10,Stone05}].} 
{In the nonlinear, coupled system of three ordinary} differential equations 
{(ODEs) for $x, y$ and $z$ below,} 
\begin{linenomath*}\begin{equation}
\label{L63}
\begin{split}
&\frac{\textrm{d}x}{\textrm{d}t} = \sigma(y-x) + \lambda_i\cos\theta_i\,,\\
&\frac{\textrm{d}y}{\textrm{d}t} = \rho x - y - x z + \lambda_i\sin\theta_i\,,\quad\frac{\textrm{d}z}{\textrm{d}t} = x y - \beta z\,
\end{split}
\end{equation}\end{linenomath*} 
{the time-constant forcing terms in the $x$- and $y$-equation represent, in fact, an addition to the forcing hidden in the original L63 model. The latter forcing is revealed by a well-known linear change of variables, in which $x$ and $y$ are left unchanged and $z \to z + \rho + \sigma$ \citep{Lor63}. In the new variables, the model of Eq.~\eqref{L63} will take the canonical form of a forced-dissipative system \cite[Sec.~5.4]{GhCh87}, with an extra forcing term $- \beta(\rho + \sigma)$ in the $z$-equation, just like the original L63 model.}

{Here $\lambda_i$ is the 
intensity of the additional forcing and $\theta_i$ is 
its} direction in world $i=0,1$: i.e., $\lambda_0=0$ 
{represents} a counterfactual world with no {additional forcing, 
while $\lambda_1\neq 0$. We take the parameters} $(\sigma,\rho,\beta)$ 
{to equal} their usual values $(10, 28, 8/3)$ {that yield the well-known chaotic behavior, and 
the (nondimensional) time unit $t$ is interpreted as equaling days.} 

{The ODE system given by \eqref{L63}} is discretized by using $\Delta t = 0.01$ 
and $t$ refers {hereafter} to the number of time increments {$\Delta t$.} 
{This system} 
is then 
{turned into one of stochastic difference equations [S$\Delta$Es: \citet{Arnold, CSG11}]} by adding an error term $\vv_t$ assumed {to be 
Gaussian and centered} with covariance $\mathbf Q = \sigma_Q^2\,\mathbf I$, where $\mathbf I$ is the $3\times 3$ identity matrix. {Furthermore, we assume 
that all} three coordinates {$(x,y,z)$} of the state vector are 
observed, 
{i.e. that $\Hs=\mathbf I$, and that} the measurement error term $\w_t$ is 
{also Gaussian and centered,} with covariance $\mathbf R = \sigma_R^2\,\mathbf I$. {Recalling the notation introduced in Sec~\ref{sec:method}a, we associate a label $\omega\in\Omega$ with each realization of the pair of random processes $(\vv_t, \w_t)$ that drive the model given by Eq.~\eqref{L63} and perturb its observations, respectively.}

The 
{S$\Delta$E system defined above} 
is stationary, i.e. the PDF of the state vector $\x_t$ depends neither on $t$ nor on $\x_0$ 
{after a sufficiently 
long time $t$. This PDF can be obtained as the (numerical) solution of the Fokker-Planck equation associated with Eq.~\eqref{L63}, and it is the mean over $\Omega$ of the sample measures obtained for each realization $\omega$ of the noises $\vv_t$ and $\w_t$ \citep[and references therein]{CSG11}. Each sample measure is supported on a random attractor that may have very fine structure and be time-dependent \citep[Figs.~1--3 and supplementary material]{CSG11}, but the PDF is supported smoothly, in the counterfactual world in which $\lambda_0 = 0$, on a ``thickened'' version of the {\mg fairly} well-known strange attractor of the original L63 model.}

{In the factual world in which $\lambda_1 \neq 0$, the nature of the PDF is quite similar, but its exact} shape is affected by the parameters $(\lambda_1,\theta_1)$ of the forcing. 
{In both worlds, the} PDFs 
can be estimated, for instance, by using kernel density estimation applied to {ensembles of} simulations obtained 
{for either forcing. In Figs.~\ref{attractor}a,b, 
we plot} the 
projections of both PDFs onto {the plane 
associated with the greatest variance in the factual PDF.} 
The difference between 
{the two PDFs is shown in Fig.~\ref{attractor}c; 
it emphasizes} the existence of an area of the state space (represented in white), which is more 
{likely} to be reached in the factual world than in the counterfactual one. 

Next, {we define} an event 
to occur for the sequence {$\{ \y_t: t = 0, \ldots, T\}$} if 
{the scalar product $\hat{\phi}' \y_{t}$ between the unit vector $\hat{\phi}$ in the direction $\phi$ and $\y_t$, i.e. the projection of $\y_t$ onto the direction $\phi$,} exceeds $u$ for some {$0 \leq t\leq T$,} where $\phi$ is a specified direction and 
$u$ is a threshold chosen based on $\phi$ so that $p_1=0.01$. {Figure~\ref{attractor}d} shows a selection of sequences from both worlds in which an event did occur, where $\phi$ was chosen 
to be the 
{leading direction in the projection 
plane.} 

For this choice of $\phi$, the 
trajectories {associated with event occurrence} happen to all 
{lie} in the area of the state space which is more 
{likely} to be reached in the factual world than in the counterfactual one. Accordingly, the probability of the event in the 
{former} is found to be higher than in the 
{latter, i.e. $p_1>p_0$,} and the occurrence of an event $\{\max_{\{0 \leq t \leq T\}} \phi'\y_t\geq u\}$ is thereby informative from a causal perspective, i.e. the associated probabilities of necessary and sufficient causation are positive. 

{Figure~\ref{attractor}d} also shows that 
{the trajectories associated with the event in the two worlds --- counterfactual (green) and factual (red) --}- 
appear to have slightly 
{distinct} features: {the red trajectories are shifted towards higher values in the second direction, of highest-but-one variance. Such distinctions might help} 
discriminate {further between the two worlds in the DADA framework.}

\subsection{{DADA for the modified L63 model}}
\label{ssec:dadaL63}

The DADA procedure is illustrated in 
Fig.~\ref{trajectories}. We plot in panel (a) a trajectory of the state vector $\x_t$ simulated under factual conditions, {i.e. in the presence of the additional forcing (black solid line), 
along with the} observations {$\{ \y_t: 0 \le t \leq T \}$ (gray dots),} with $T=400$. The EnKF is used to assimilate these observations into a factual model {($i = 1$)} that thus matches the true model {\mg $\M = \M_1 = \M(\lambda_1, \theta_1)$}
used for the simulation: a reconstructed {trajectory is obtained from the 
{\mg corresponding} analyses 
 $\x^a_t$ (red solid line in panel (a)), cf. Eqs.~\eqref{KF}, 
 and the likelihoods $f_1(\y_t)$ (red solid line in panel (c))} are obtained by application of Eq.~\eqref{llkKF}, respectively.
 
Next, the assimilation is repeated 
in the counterfactual model {($i = 0$, i.e. $\lambda = 0$)} to obtain a second analysis of the 
{trajectory, from the same observations; see green solid line in panel (a), for $T = 400$. 
The corresponding likelihoods $f_0(\y_t)$ are shown in panel (c) {\mg as a green solid line.} Comparing the 
trajectories of the two analyses 
{in Fig.~\ref{trajectories}a shows} that, even though the counterfactual analysis 
{(green line) uses the same data as the} factual analysis (red line), 
the former lies 
closer to the true trajectory {\mg (black line). 

The} local discrepancies between the 
{\mg trajectories estimated in the two worlds} appear to be rather small at first glance, 
{\mg cf. panel (a), and so are the instantaneous differences between the associated factors on the right-hand side of Eq.~\eqref{llkKF}; the latter are shown as gray rectangles in panel (c) of the figure.
Still, the evidence in favor of the factual world accumulates as the time $t$ over which the two trajectories 
differ, albeit by a small amount, lengthens.}
This cumulative difference {\mg in evidence, $\log f_0(\y_t) - \log f_1(\y_t)$,} is reflected by a 
{\mg growing} gap between 
the two {\mg curves, red and green, in panel (c), and by} an associated high {\mg mean growth over time of the probability PN of necessary causation, 
cf. the black solid line in panel (d).}

In order to evaluate more systematically its performance and robustness {\mg compared} 
to the conventional {FAR} approach, the DADA procedure was applied to a large sample of sequences {$\y_t$} of length $T=20$ simulated under diverse conditions. The sample explored all possible combinations of {the triplet of parameters $(\lambda_1, \sigma_Q, \sigma_R)$, with ten equidistributed} values each,
{for a total of} $10^3$ combinations; 
{the ranges were $0 \leq \lambda_1 \leq 40$, $0.1 \leq \sigma_Q \leq 0.5$ and $0.1 \leq \sigma_R \leq 1.0$, respectively,} with $\theta_1=-140^\circ$. For each combination {of $(\lambda_1, \sigma_Q, \sigma_R)$,} ten directions $\phi$ were randomly generated and $u$ was defined based on $\phi$ as {in Sec.~\ref{sec:L63}a 
above, so as to achieve $p_1 \geq 0.01$.}  

In order to estimate the corresponding conventional probabilities $p_0$ and $p_1$ of the associated event 
{defined as $\{\max_{\{0 \leq t \leq T\}} \phi'\y_t\geq u\}$, $n=50~000$} sequences {$\y_t$ of length $T = 20$} were simulated, 
{by using} a single sequence of length $nT=10^6$ 
and 
{splitting it into $n$ equal segments.} Probabilities $p_0$ and $p_1$ were then directly estimated from empirical frequencies because the high value of $n$ here did not require the use of the EVT extrapolation normally used for smaller $n$. 

For each 
{quintuplet of parameter values 

\noindent$(\lambda_1,\sigma_Q,\sigma_R; \phi,u)$,} one hundred sequences of observations {$\{ \y_t: 0, \ldots, T = 20 \}$} 
were generated with a proportion $p_1/(p_1+p_0)$ being simulated from the factual world and a proportion $p_0/(p_1+p_0)$ from the counterfactual one. All sequences were treated with the DADA procedure {--- by applying DA to the synthetic observations according to Eqs.~\eqref{KF1}--\eqref{KF4} --- and then Eq.~\eqref{llkKF}
} to obtain $f_0(\y)$ and $f_1(\y)$ {from the reconstructed trajectories.} The 
a priori mean and covariance $\x^f_0$ and $\mathbf P_0^f$ required as inputs to the DADA procedure were those associated {with} the PDF of the attractor, 
{given the forcing conditions $(\lambda_1 \in [0, 40], \theta_1 = - 140^\circ)$
assumed for 
each assimilation experiment.} 
As a result, two probabilities 
{PN of necessity} are finally obtained for each sequence {$\y_t$, ${\rm PN}_p = 1-p_0/p_1$ for the conventional approach and ${\rm PN}_f = 1-f_0(\y)/f_1(\y)$ for the DADA approach.}  

We next wish to evaluate under various conditions how well the two probabilities 
{${\rm PN}_p$ and ${\rm PN}_f$} perform 
{with respect to} discriminating between the factual and counterfactual forcings. Consider a simple discrimination rule whereby {a trajectory $\y_t$} is identified as factual for PN exceeding a given threshold, and as counterfactual otherwise. The so-called {receiver operating characteristic (ROC)} 
curve plots the rate of true positives as a function of the rate of false positives obtained when varying the threshold {in a binary classification scheme} from 0 to 1; it thus gives an overall visual representation of 
{the skill of our PN} as a discriminative \textit{score}. 

The 
{\cite{Gini21} index $G$ was originally introduced as a measure of statistical dispersion intended to summarize the information contained in the \cite{Lor05} curve that represents the income distribution of a nation's residents; $G$} may be viewed, {though, more generally} as a metric summarizing {the dispersion of any smooth curve that starts at the origin and ends at the point $(1,1)$ with respect to the diagonal of the corresponding square. In particular, we use $G$ here to summarize into a single scalar} the ROC curve, 
which ranges from 0 for random discrimination to 1 for perfect discrimination. 

{Figure~\ref{time}a} shows ROC curves obtained over the entire sample of $n = 50~000$ sequences: 
they correspond to $G = 0.35$ for the conventional method and to $G = 0.82$ for the DADA method, i.e. the {\mg overall} performance gap is more than twofold. 
As expected, the performance of both methods is nil for $\lambda_1=0$ and {\mg it} is very sensitive to the intensity of the forcing, {\mg cf. 
Fig.~\ref{time}b.

Furthermore,} the skill of the DADA method is boosted when decreasing the level of model error, 
{\mg cf. Fig.~\ref{time}c; this is an expected result, since} 
DA becomes more reliable when the model is more accurate, {\mg and when it} is known to be so. Ultimately, under perfect model conditions, {\mg i.e. as $\sigma_Q\to0$,} DADA reaches perfect discriminative power, 
with G$\to1$, no matter how small, but {\mg still} positive, the forcing is; 
{\mg see Fig.~\ref{time}d.} On the other hand, the level of observational error $\sigma_R$ appears to have {\mg but a} limited effect on {\mg DADA} performance for the range of values considered, 
{\mg cf. Fig.~\ref{time}e.}

Finally, {\mg Fig.~\ref{time}f shows that} both methods perform better when the contrast between $p_0$ and $p_1$ is strong, 
but the latter does not influence the gap between the two methods, which remains nearly constant. This constant gap thus appears to quantify the additional power resulting from the extra discriminative features that the PDF $f(\y)$ is able to capture on top of those associated 
{\mg with} the probability $P(\phi(\y)\geq u)$.


\section{Discussion and {\mg conclusions}}
\label{sec:concl}

{\mg \cite{HPONG14} have relied on the} causality theory of \cite{Pearl00} 
{\mg to show} that the ratio between the factual evidence $f_1(\y)$ and the counterfactual 
{\mg evidence} $f_0(\y)$ is 
{\mg important in studying} causal attribution of weather- and climate-related events. 
{\mg In this paper, we first described data assimilation (DA)} methods 
{\mg and then demonstrated that they are 
well suited for} deriving 
{\mg $f_0(\y)$ and $f_1(\y)$ from trajectories in the factual and the counterfactual worlds, respectively. Besides, 
these methods} offer the key practical advantage 
{\mg of being} already up-and-running in near real time at meteorological centers. 

Combining these two sets of considerations, {\mg theoretical and practical, opens} a novel route 
towards near real time, systematic causal attribution of weather- and climate-related events, thereby addressing 
{\mg a key} challenge in {\mg the field of detection and attribution (D\&A)} at present \citep{Stott13}.
 
\subsection{{\mg Theoretical considerations}}
\label{ssec:theory}

Implementing the 
{\mg DA for D\&A (DADA)} approach in the context of the L63 model {\mg in Section~\ref{sec:L63}} allowed for a detailed step-by-step illustration of our methodological proposal. It also provided a basic test 
for an initial performance assessment, which showed an improved level of {\mg discriminating power 
with respect to} the conventional approach {\mg outlined in Section~\ref{sec:intro}.} 
{\mg These} results are promising, 
{\mg and their promise is easy to understand, given the fact that} the DADA approach 
leverages the 
available information {\mg on the entire trajectory} $\y$, as opposed to the single specific feature $\mathbf{1}_{\phi(\y)\geq u}$ in the conventional approach.

It is 
important, {\mg though, to stress}  
that the term ``performance'' here should be considered with caution: 
improving 
{\mg discriminatory} performance may or may not be a desirable outcome, depending on the causal question being asked. 
{\mg \cite{HPONG14} have shown that the causal question being formulated} reflects the subjective interests of a particular class of end-users, and 
{\mg that the formulation itself may} 
dramatically affect the answer. 

For example, the question ``\textit{did anthropogenic {\mg CO$_2$} emissions cause the heatwave observed over Argentina during January 2014?}'' {\mg has been traditionally treated by defining a 
``heatwave'' 
in terms of a} predefined temperature index reaching a predefined threshold, 
{\mg i.e., 
by} a singular index 
{\mg exceeding} a singular threshold. This class of questions matters for instance in the context of insurance {\mg disbursements, 
where} a financial compensation may typically be triggered based on such an index exceedance. In this situation, the additional 
{\mg discriminatory} power of DADA is meaningless because the DADA computation 
does not address the question at stake: there is {\mg simply} no alternative to 
{\mg computing} the probabilities $p_0$ and $p_1$ of the index exceeding the threshold. 

However, if the question is {\mg formulated instead as} ``\textit{did anthropogenic 
{\mg CO$_2$} emissions cause the atmospheric conditions observed over Argentina during January 2014?}'' {\mg --- i.e.,} without specifying which feature of the observed sequence is most important --- then improving discrimination makes 
{\mg perfect} sense and DADA becomes fully relevant. {\mg Furthermore,} DADA 
{\mg is still} fully relevant even if the question is 
{\mg formulated more specifically as} 
``\textit{did anthropogenic 
{\mg CO$_2$} emissions cause the damages generated in Argentina by the atmospheric conditions of January 2014?,}'' provided {\mg that is, that} a model relating atmospheric observations to damages at every 
time step $t$ {\mg along the trajectory of the physical model used in the assimilation 
is available and 
can be} integrated into the 
{\mg observation} operator $\Hs$.

On the other hand, the results of 
{\mg Section~\ref{sec:L63}} should also be considered with caution simply because the L63 testbed obviously differs in many {\mg respects} from the real situation envisioned for future applications, both in terms of model dimension 
{\mg $n$} and observation dimension 
{\mg $d$: in practice $n$ will be very large and $d \ll n$, while here we took $d = n =3$.}

In particular, 
choosing a 
{\mg highly idealized,} climatological a priori distribution on the initial condition $\pi(\x_0)$ does not raise any difficulty under the tested conditions nor {\mg does it influence} significantly the outcome of the procedure (not shown). 
The choice of $\pi(\x_0)$, however,} may be an important problem in practice, 
{\mg when $d\ll n$, and lead} to potentially spurious results. 

As a consequence, it may be 
{\mg both necessary and useful} to further constrain the {\mg so-called {\em background PDF} $\pi(\x_0)$}
by using the {\mg forecasts} originating from $\tau$ previous assimilation cycles, {\mg thus following the ideas of lagged-averaged forecasting \citep{LAF83, Dalcher}. 
The} evidence thus obtained, {\mg though, will then also depend on} previous observations over the ``initialization'' window $[-\tau,...,-1]$ --- i.e., it {\mg will no longer represent exclusively} the desired evidence $f(\y)$. 
Besides, choosing $\tau$ optimally to constrain the initial background {\mg PDF in a satisfactory manner,} while at the same time limiting the latter unwanted dependence {\mg on} previous observations, is a challenging question that 
{\mg needs} to be adressed. 

More generally, the problem of evaluating the evidence $f(\y)$ is not new in {\mg the HMM and} DA literature; 
{\mg see, for instance,} \cite{Baum70,HK01,Pitt02} {\mg and \cite{Kan09}.} 
{\mg Various algorithms are thus} available to 
{\mg carry out this evaluation,} depending on a number of key assumptions 
{\mg --- such as lack of Gaussianity or linearity ---} and on the inferential setting chosen, e.g. particle filtering. These {\mg algorithms} may provide 
{\mg accurate and effective} solutions to the above problem, as well as improved alternatives to the Gaussian {\mg and linear} approximation of {\mg Eq.~\eqref{llkKF}, since the latter may 
not be sufficiently accurate} for succesfully implementing the DADA approach under realistic conditions.





\subsection{{\mg Practical considerations}}
\label{ssec:practice}

While 
{\mg we have shown here that the proposal of using 
{\mg DADA for event attributions} has intellectual merit, its} 
main strength 
{\mg lies, in our view, in} down-to-earth cost considerations. By design, 
the DADA approach 
allows {\mg one} to piggyback at a low marginal cost on the 
large and powerful infrastructures already in place at {\mg several} meteorological centers, 
{\mg in terms of both hardware and personnel. These centers} are capable of processing 
{\mg massive} amounts of observational data with high-throughput pipelines on the world's largest computational platforms, as opposed to requiring the design, set-up and maintenance of a new and large, D\&A-specific infrastructure {\mg to collect} observations and {\mg generate ---} under near real time constraints 
--- the many model simulations 
{\mg required} by the conventional approach recalled {\mg in Section~\ref{sec:intro}.} 

Taking a step back, it is useful to examine our proposal within the wider context of the emergence of so-called climate services. It is 
{\mg widely} recognized that extending the scope of activity of meteorological centers from {\mg being} ``monoline'' weather forecasting providers to {\mg becoming} ``multiline'' climate services providers --– encompassing, {\mg for instance,} weather {\mg forecasting} and weather event attribution as two service lines among several others --€" is a relevant strategic option \citep{Hew12}. 
{\mg Such a strategy} may foster the timely and cost-efficient emergence of the latter services by building upon technological and infrastructure synergies with the former. 
For {\mg these} 
reasons, our proposal is particularly relevant for, and could contribute to, the implementation of {\mg the strategic option just outlined. 

This being} said, DADA 
{\mg can very well serve} as a method for near real time event attribution even for hypothetical 
climate services providers 
{\mg that focus uniquely or mainly} on longer time scales, 
{\mg beyond a month,   a season or a year.} In such a context, DADA may 
allow for the assimilation of 
{\mg a broader range of} observations, and in particular of ocean observations; 
 {\mg it may, in fact, be important to include} the latter in 
 causal analysis when the event occurrence under scrutiny is defined over a {\mg sufficiently large} time 
 window.


\begin{acknowledgements}
This work has been supported by grant DADA {from the Agence Nationale de la Recherche (ANR, France: AH and all co-authors) and by the Multi-University Research Initiative (MURI) grant N00014-12-1-0911 from the the U.S. Office of Naval Research (MG).}
\end{acknowledgements}




\textbf{Appendix: Derivation of the model evidence}

In this appendix, we 
{\mg outline} 
the derivation of model evidence  within 
{\mg a} general Bayesian framework, and we apply the latter to the 
{\mg narrower} KF context to obtain Eq.~\eqref{llkKF}. Consider two consecutive cycles of a DA run, the first with state vector $\x_t$
and observation vector $\y_t$ at instant $t$ and the subsequent one with state vector $\x_{t+1}$
and observation vector $\y_{t+1}$ at instant $t+1$. We plan to find a tractable expression for the model evidence
$p(\y_t, \y_{t+1})$.

The model evidence 
{\mg provided by} the full sequence of observations $\y=(\y_0,...,\y_T)$ will be inferred by recursion,
using the results of this {\mg two-observation} setting. In order to decouple the two cycles, 
{\mg one first 
has to spell} out the 
{\mg Bayesian inference} $p(\y_t, \y_{t+1}) = p(\y_t) p(\y_{t+1} | \y_t)$. We look for a tractable expression for $p(\y_{t+1} | \y_t)$ by further introducing the states $\x_{t+1}$ and $\x_t$ as 
{\mg intermediate}
random variables:
\begin{linenomath*}\begin{equation}
\label{decomposition}
{\mg 
\begin{split}
p(\y_{t+1} | \y_t) =& \int_{\x_{t+1}} \! p(\y_{t+1} | \y_t, \x_{t+1}) p(\x_{t+1} | \y_t)\,{\rm d}\x_{t+1} \\
= &\int_{\x_{t+1}} \! p(\y_{t+1} | \x_{t+1})\\
&\times\left\{\int_{\x_t} \! p(\x_{t+1} | \x_t)\,p(\x_t | \y_t)\,{\rm d}\x_t\right\} {\rm d}\x_{t+1}\,,
\end{split}
}
\end{equation}\end{linenomath*}
where $p(\y_{t+1} | \x_{t+1})$ is the likelihood of the observation vector $\y_{t+1}$ conditional on the state vector $\x_{t+1}$ and it is known from 
{\mg Eq.~\eqref{obs}.} 

{\mg The 
conditional PDF $p(\x_t | \y_t)$} of $\x_t$ on $\y_t$ at instant $t$ {\mg --- 
which appears on the right-hand side of the above equation ---}
is referred to as the {\mg {\em analysis}} PDF in {\mg the DA literature, where it is denoted by a 
superscript $a$ \citep{Ide97}, and it} constitutes 
{\mg the main DA} output. 
The integral $\int_{\x_t} \! p(\x_{t+1} | \x_t) p(\x_t | \y_t)\,{\rm d}\x_t=p(\x_{t+1} | \y_t)$, in which $p(\x_{t+1} | \x_t)$ is known from the model dynamics given by 
{\mg Eq.~\eqref{dyn},} 
propagates this analysis PDF 
{\mg further in time, to} instant $t+1$. Hence, 
{\mg the result of this integration} coincides with the forecast PDF, 
{\mg denoted by superscript $f$ in the DA literature \citep{Ide97}. 
It follows that}
this decomposition is tractable using a DA scheme 
{\mg that} is able to estimate the conditional and forecast PDFs.

Next, let us apply 
{\mg the general Bayesian inference \eqref{decomposition}} to the case 
{\mg in which all the PDFs involved} are Gaussian; 
{\mg this} requires, {\mg in turn, that both the 
dynamics} and observation models {\mg $\M$ and $\Hs$} be linear, and that the input statistics all be Gaussian. In 
{\mg this} case, the Kalman filter allows for the exact computation of the PDFs mentioned in 
{\mg Eq.~\eqref{decomposition},}
which turn out to be Gaussian. 

In the following, ${\cal N}(\overline{\x},\bP)$ designates the Gaussian PDF of mean $\overline{\x}$ and covariance matrix $\bP$. 
In this context, the analysis PDF at instant $t$ is ${\cal N}(\x^a_t,\bP^a_t)$, where $\x^a_t$ and $\bP^a_t$ are the analysis state and error covariance matrix at instant $t$. As a result of {\mg the linearity 
assumptions,} the forecast PDF at instant $t+1$ is given by a Gaussian distribution ${\cal N}(\x^f_{t+1},\bP^f_{t+1})$, where $\x^f_{t+1}$ and $\bP^f_{t+1}$ are the forecast state and error covariance matrix at instant $t+1$. 
Further, the integration on $\x_{t+1}$ in Eq.~(\ref{decomposition}) can readily be performed under these circumstances, with the outcome that $p(\y_{t+1} | \y_{t})$ is distributed as ${\cal N}(\bH\,\x^f_{t+1},\bR+\bH\,\bP^f_{t+1}\,\bH')$. 

The desired model evidence $f(\y)$ can
then be computed by recursion on successive time steps as:
\begin{linenomath*}\begin{equation}
\label{full}
\begin{split}
&f(\y) = p(\y_0)\,\prod_{t=1}^{T} (2\pi)^{-\frac{d}{2}} |\bSigma_t|^{-\frac{1}{2}}\\
&\times\exp
\left\{
-\frac{1}{2}(\y_t-\bH\x^f_t)'\bSigma_t^{-1}(\y_t-\bH\x^f_t)
\right\}\,;
\end{split}
\end{equation}\end{linenomath*}
{\mg here} $p(\y_0)$ represents the {\mg prior PDF of}  
the initial 
{\mg state,} $\bSigma_t = \bR+\bH\bP^f_t\bH'$, and 
This expression coincides with {\mg Eq.~\eqref{llkKF}} and can be 
{\mg evaluated} with the help of any DA method that yields the forecast states and forecast error covariance matrices, such as the KF or the EnKF. Note that the traditional standard Kalman smoother would give the same result as the KF, since they share the same forecasts.

Finally, {\mg Eqs.~\eqref{decomposition} and \eqref{full}} above show that the likelihood $f(\y)$ may be obtained as a by-product of the inference on the state vector $\x$, which usually is the main purpose in 
{\mg numerical weather prediction.} This idea may actually be highlighted {\mg in even 
greater generality} by considering the 
equality: 
\begin{linenomath*}\begin{equation}
\label{bayes2}
 {f(\y) = \frac{p(\y \mid \x)
 p(\x)}{p(\x\mid\y)}\,.}
\end{equation}\end{linenomath*}
{\mg While Eq.~\eqref{bayes2} is} a direct consequence of Bayes theorem, 
{\mg it also} illustrates a point 
{\mg that} is arguably not so intuitive. The likelihood {\mg $f(\y)$} is 
obtained here as the ratio of two quantities: 
a numerator 
{\mg $p(\y \mid \x)p(\x)$ that} is a model premise inherently postulated by {\mg Eqs.~\eqref{obs} and \eqref{dyn}, and a denominator $p(\x\mid\y)$ 
{\mg that} may be viewed as the end result of the primary inference on $\x$.} 
In other words, 
{\mg estimating $f(\y)$ 
requires only a straightforward division,} provided $\x$ has been previously inferred. 

Equation~\eqref{bayes2} thus expresses with great clarity and simplicity a fundamental idea buttressing our proposal, {\mg as it provides} a general theoretical justification {\mg for the  
suggestion} 
{\mg of deriving the} likelihood 
from 
{\mg an} inferential treatment {\mg that focuses} on $\x$. To put it 
{succintly, this equation basically 
says, ``\textit{He {\mg who can do more can do less.}''} {\mg In the context 
of DA, whose} end purpose is to infer the state vector $\x$ out of an observation $\y$ --- i.e., the \textit{more} part --- it is possible to obtain the likelihood as a by-product thereof --- i.e., the \textit{less} part --- and thus almost for free.

\clearpage


\begin{figure} 
\begin{center} 
\includegraphics[angle=0, width=13cm]{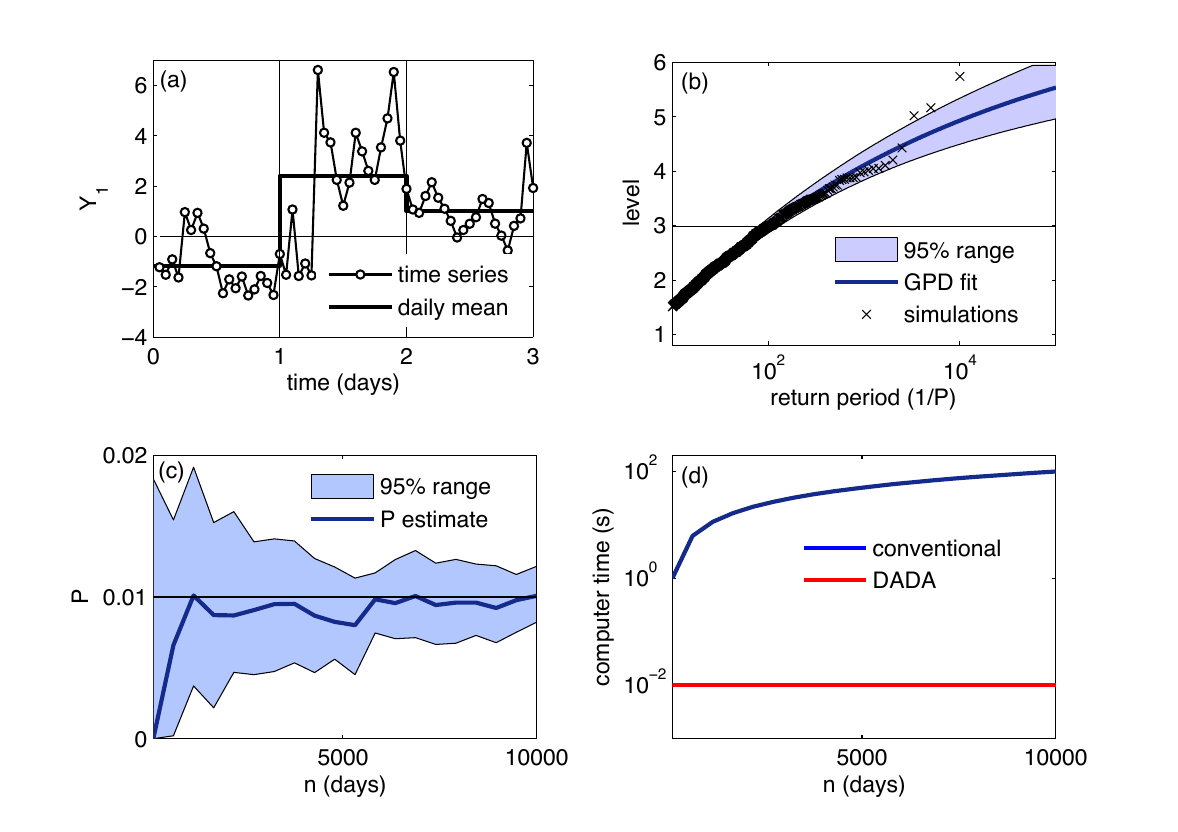} 
\end{center}
\caption{Illustration {of the conventional D\&A approach as applied to a univariate} AR(1) process. (a) Observed time series (first component $Y_1$, dotted line) and daily average $\phi(\Y)$ 
{(heavy solid} line). (b) Threshold level (vertical axis) as a function of the return period (horizontal axis): simulated values (crosses); fit based on the Generalized Pareto distribution (GPD, {heavy dark-blue} line); uncertainty range {at the 95\% level} (light blue area); {and} threshold value $u=3.1$ (light {solid} black line). (c) Estimated value of $P=P(\phi(\Y)\geq u)$ ({heavy dark-blue} line) using a GPD fit as a function of the sample size $n$ (horizontal axis); uncertainty range (light blue area); {and} true value $P=0.01$ (light {solid} black line). (d) Computational time on a desktop computer (seconds, vertical axis) as a function of sample size $n$ (horizontal axis) required by the conventional method (dark blue line) and the DADA method ({solid} red line); {the latter method is explained in Sections~\ref{sec:method}b and \ref{sec:L63} below.} }
\label{pdfex}
\end{figure}

\clearpage
\begin{figure} 
\begin{center} 
\includegraphics[angle=0, width=14cm]{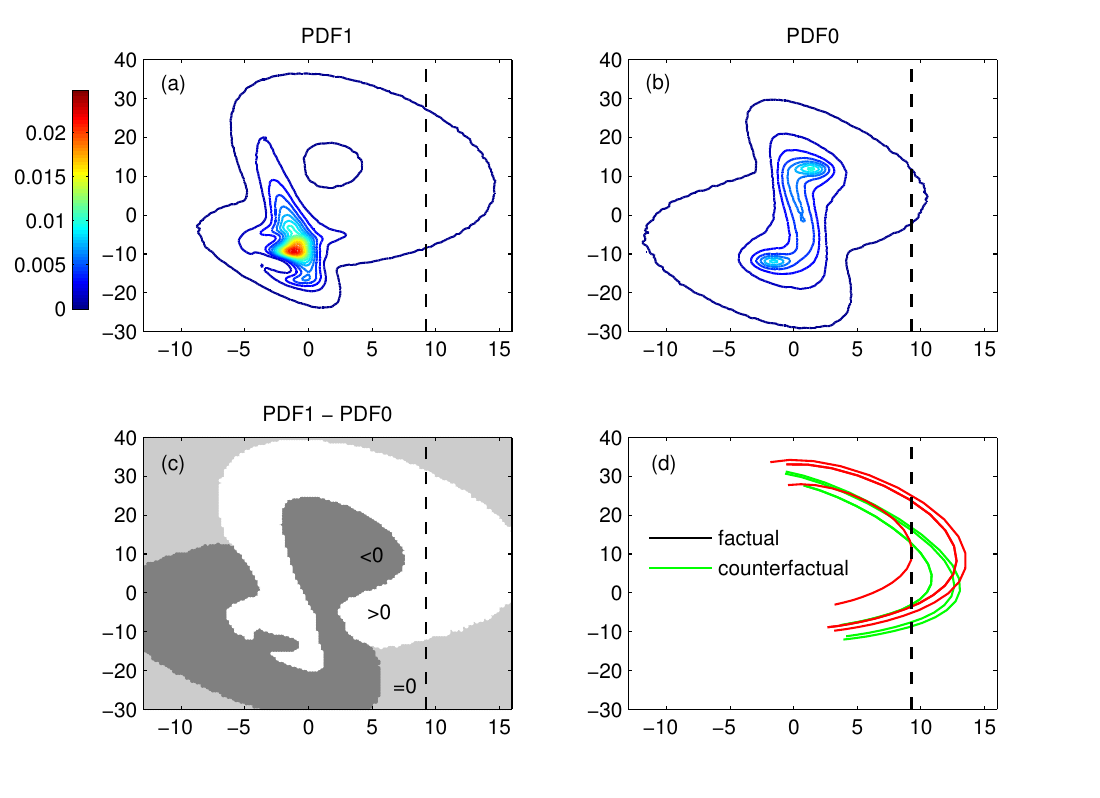} 
\end{center}
\caption{Two-dimensional (2-D) projections of the PDF of the modified L63 model; the projection is onto a plane defined by the two leading eigenvectors of the factual PDF shown in the first panel. 
(a) PDF of the factual attractor, {with $\lambda_1=20$ and $\sigma_Q=0.1$; and} (b) PDF of the counterfactual attractor, {with $\lambda_0 = 0$.} (c) Difference between the factual and counterfactual PDFs. (d) Sample 
trajectories 
{associated with} an event occurrence originating from the factual (red {solid} lines) and counterfactual worlds (green {solid  lines); the vertical dashed line in all four panels indicates the threshold $u$ 
with respect to the horizontal axis of largest variance in the factual PDF.}}
\label{attractor}
\end{figure}

\clearpage
\begin{figure} 
\begin{center} 
\includegraphics[angle=0, width=12.5cm]{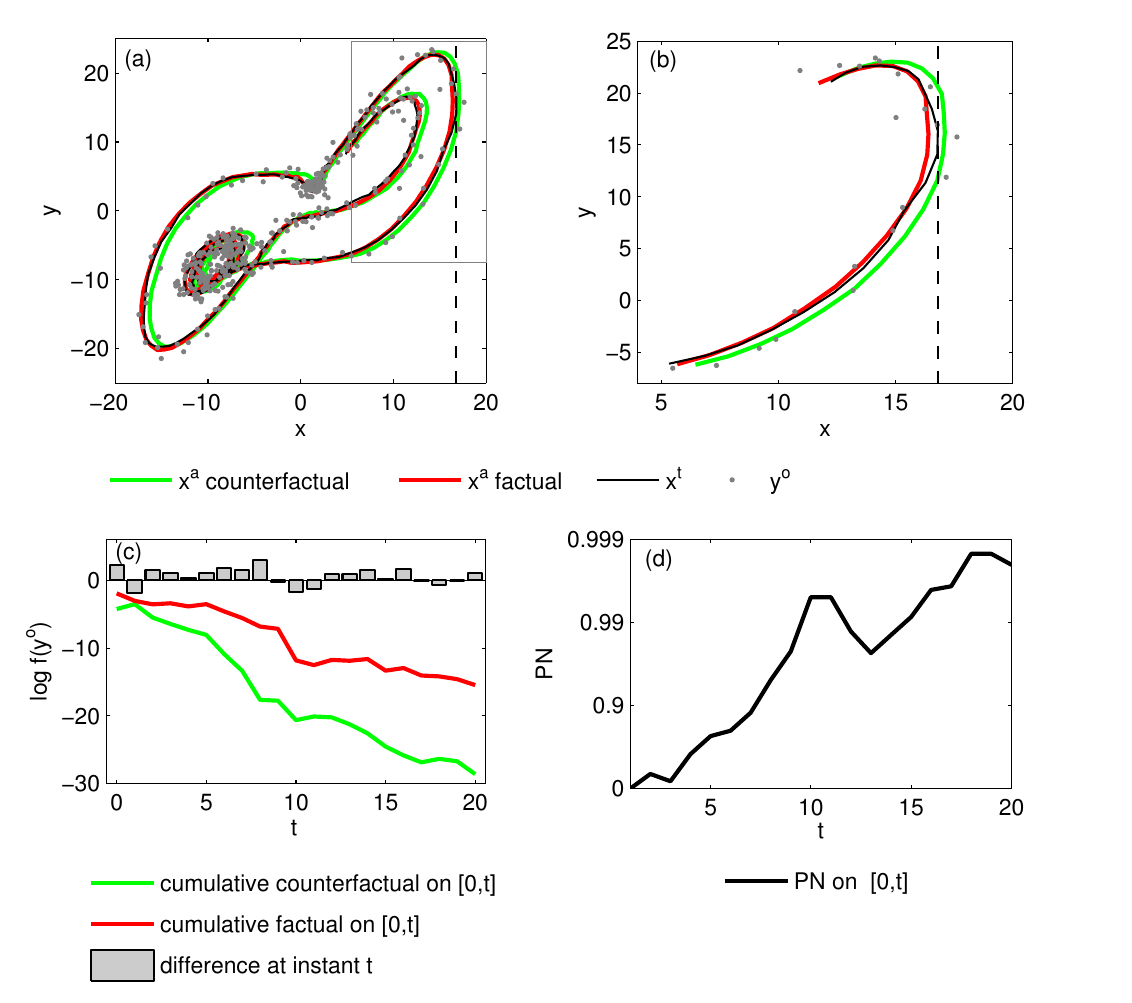}
\end{center}
\caption{Sample trajectories from data assimilation (DA) in our modified L63 model. (a) True trajectory 
(black {solid} line) and 
{the two} trajectories {reconstructed by DA} in the factual {($i = 1$) and counterfactual ($i = 0$)} worlds (red and green {solid lines), respectively,} 
over a long sequence, $T = 400$; the values of $\lambda_1$ and $\theta_1$ here are the same as in Fig.~\ref{attractor}, and the
assimilated observations are shown as gray dots. (b) Same as {panel (a) but} zoomed over a short sequence, {$T = 20$.} (c) Logarithm of the {\mg cumulative} evidences $f_1(\y)$ and $f_0(\y)$ (red and green lines, respectively) computed over the window {\mg $[0, t \le T]$; gray bars 
indicate} the instantaneous differences between $f_1(\y_t)$ and $f_0(\y_t)$. (d) PN computed over the window $[0,t]$.}
\label{trajectories}
\end{figure}

\clearpage
\begin{figure} 
\begin{center} 
\includegraphics[angle=0, width=13cm]{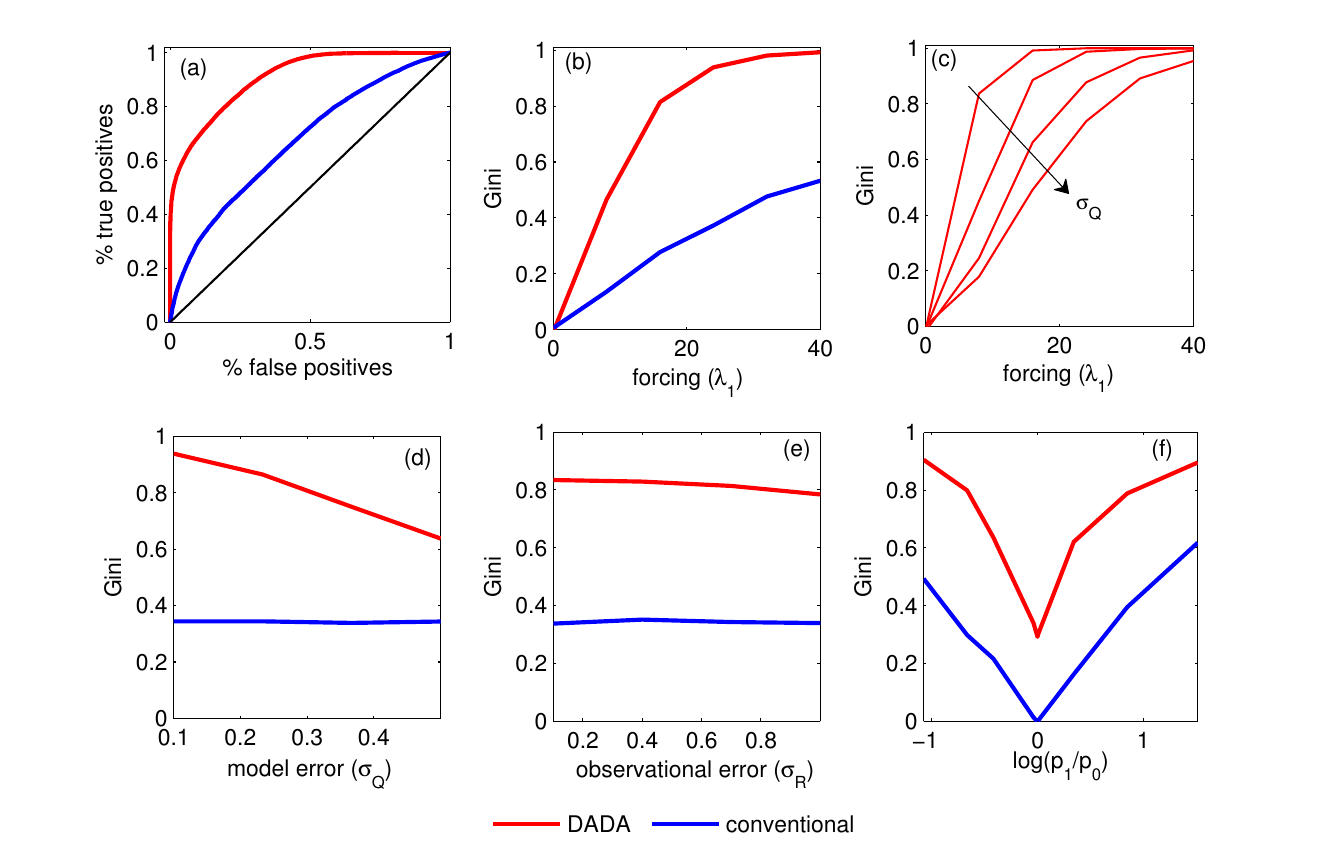}  
\end{center}
\caption{Performance of the DADA and conventional methods (red {vs.~blue solid lines, respectively).} (a) {Receiver operating characteristic (ROC)} 
curve: true positive rate 
as a function of false positive rate, 
when varying the cut-off level {$u$, as} obtained from the entire sample {of $n = 50~000$ sequences; see text for details.}. (b) Gini index 
{$G$} as a function of forcing intensity 
{$\lambda_1$.} 
(c) Same as (b) for several values of 
{$\sigma_Q$} and for DADA only, {with the black arrow indicating the direction of growing $\sigma_Q$.}  (d) Same as (b) but as a function of model error amplitude {$\sigma_Q$.} (e) Same as (b) but as a function of observational error amplitude {$\sigma_R$.} 
(f) Same as (b) as a function of the {logarithmic} contrast between the conventional probabilities $\log p_1/p_0$.
}
\label{time}
\end{figure}

\clearpage

\end{document}